\documentclass[prb,reprint,a4paper,floatfix,superscriptaddress]{revtex4-1}

\usepackage{graphicx}
\usepackage{amsmath}
\usepackage{hyperref}
\usepackage{xfrac}
\usepackage{xcolor}

\newcommand{\tcm}{T_{\mathrm{c}} }

\newcommand{\tcmaxm}{ T _{\mathrm{c}} ^{\mathrm{max}} }
\newcommand{\fig}{Fig.}

\newcommand{\ekm}{\epsilon(\mathbf{k}) }
\newcommand{\ekpm}{\epsilon(\mathbf{k}^{\prime}) }
\newcommand{\kk}{\mathbf{k} }

\begin{document}

\title{Possible very high transition temperatures in the infinite-layer ACuO$ _{2} $ cuprate superconductor for A=\{Mg, Ca, Sr, Ba\}: A DFT study}

\author{B.P.P. Mallett}
\email{benjamin.mallett@gmail.com}
\affiliation{Robinson Research Institute, Victoria University, P.O. Box 600, Wellington, New Zealand}
\affiliation{MacDiarmid Institute for Advanced Materials and Nanotechnology, Victoria University of Wellington, P.O. Box 600, Wellington 6140, New Zealand}

\author{N. Gaston}
\affiliation{MacDiarmid Institute for Advanced Materials and Nanotechnology, Victoria University of Wellington, P.O. Box 600, Wellington 6140, New Zealand}
\affiliation{MacDiarmid Institute for Advanced Materials and Nanotechnology, Department of Physics, The University of Auckland, Auckland, New Zealand}

\author{J.G. Storey}
\affiliation{Robinson Research Institute, Victoria University, P.O. Box 600, Wellington, New Zealand}
\affiliation{MacDiarmid Institute for Advanced Materials and Nanotechnology, Victoria University of Wellington, P.O. Box 600, Wellington 6140, New Zealand}

\author{G.V.M. Williams}
\affiliation{MacDiarmid Institute for Advanced Materials and Nanotechnology, Victoria University of Wellington, P.O. Box 600, Wellington 6140, New Zealand}

\author{A.B. Kaiser}
\affiliation{MacDiarmid Institute for Advanced Materials and Nanotechnology, Victoria University of Wellington, P.O. Box 600, Wellington 6140, New Zealand}

\author{J.L. Tallon}
\affiliation{Robinson Research Institute, Victoria University, P.O. Box 600, Wellington, New Zealand}

\date{\today}

\begin{abstract}

We show from a bond valence sum correlation that very high superconducting $ T_{c} $ values should be found in optimally hole-doped infinite-layer ACuO$ _{2} $ cuprates - up to 160~K for A = Ba. The projected increase in $ T_{c} $ across the series arises from ``internal pressure'' effects as A runs from Mg to Ba. We then use density functional theory to investigate these pressure effects on the band structure in an attempt to understand this progressive increase in $ T_{c} $. Where these materials have been synthesised we find good agreement between our calculated structural parameters and the experimental ones. We find that internal pressure associated with increasing ion size does indeed enhance the superconducting energy gap, as observed, via modifications to the electronic dispersion. Furthermore, in our calculations, pressure alters the dispersion independently of how it is applied (internal or external) so that the superconducting energy gap correlates with the unit-cell volume and a Fermi-surface shape-parameter describing ratio of next-nearest-neighbor to nearest-neighbor hopping integrals. We infer an energy scale for the pairing interaction of the order of 1~eV, well above the magnetic energy scale.

\end{abstract}

\maketitle

\section{introduction}
Ion-size is a concept for the volume of space occupied by an ion's bound electrons. It is an important consideration in the synthesis of novel materials or in predicting structural stability and can play an analogous role to external-pressure by altering orbital overlap.  The associated ``effective pressure'' is called chemical- or internal-pressure.\cite{marezio2000, khosroabadi2004}   Here, we refer to the chemical-pressure from altering ion-size as `internal-pressure' and mechanically applied pressure as `external-pressure'.  Isovalent ion-substitution can thus be utilized to alter the structural and electronic properties of a material, and this can be achieved in a systematic manner where the ion-size progressively changes, as in the rare-earth series R=\{Lu, \ldots, La\}, or column II ions A=\{Mg,Ca,Sr,Ba,Ra\}.  For example, in the R(Ba$ _{2-x} $Sr$ _{x} $)Cu$_3$O$_7$ family of cuprate high-temperature superconductors the lattice and structural parameters systematically change with ion-size, demonstrating internal-pressure effects.\cite{guillaume1994,licci1998, gilioli2000} Concurrently, as the ion-size is increased, the maximum superconducting transition temperature, $ \tcmaxm $, increases from 70~K for YSr$ _{2} $Cu$ _{3} $O$ _{7-\delta} $,\cite{gilioli2000} to nearly 100~K for LaBa$ _{2} $Cu$ _{3} $O$ _{7} $.\cite{lindemer1994}

\begin{figure}
	\includegraphics[width=0.95\columnwidth]{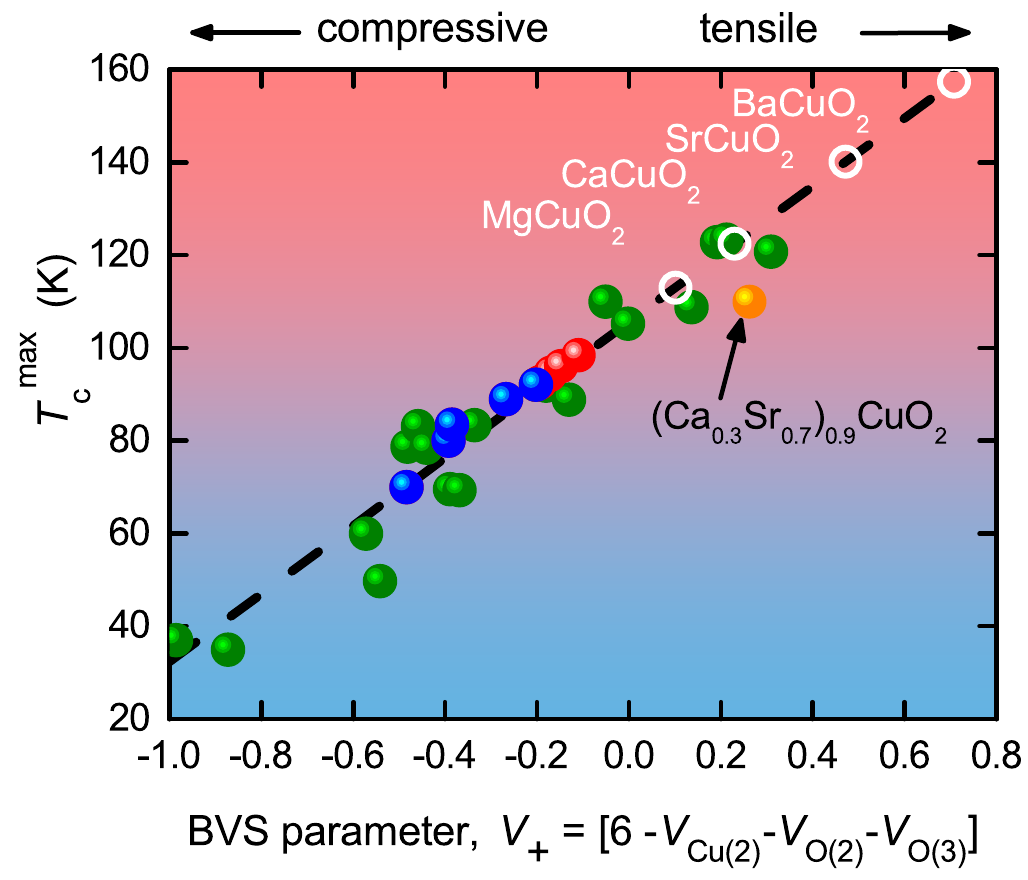}
	\caption{\label{fig:0}(Color online) The correlation between the highest $ \tcm $, $ \tcmaxm $, across all cuprates and the composite bond-valence sum $ V_{+} $. The hollow symbols show predicted $ \tcmaxm $ for optimally doped ACuO$ _{2} $ and Nd$ _{2} $CuO$ _{4} $ based on their calculated $ V_{+} $. The (Ca$ _{0.3} $Sr$ _{0.7} $)$ _{0.9} $CuO$ _{2} $ sample is likely underdoped.}
\end{figure}

Relating $ \tcm $ in a systematic way to crystallographic and electronic structure is undoubtedly complicated by the competing correlations that exist in the cuprates.\cite{keimer2015} The presence of a pseudogap as a separate competing entity, charge-ordering and their respective Fermi-surface reconstructions, is now well established.\cite{tallon2001,comin2016} Despite this, for all cuprates the maximum superconducting transition temperature, $ \tcmaxm $, in the phase curve at optimal doping does correlate exceptionally well with a structural bond valence sum (BVS) parameter, $ V_{+} = 6 - V_{\mathrm{Cu}} - V_{\mathrm{O(2)}} - V_{\mathrm{O(3)}} $, which combines both the planar copper and oxygen BVS as a measure of the distribution of doped charge between copper and oxygen orbitals.\cite{tallon1990} $ V_{+} $ also reflects the in-plane stretch and the displacement of the apical oxygen. Its correlation with $ \tcmaxm $ is shown in Fig.~\ref{fig:0} for the above-mentioned RBa$ _{2-x} $Sr$ _{x} $Cu$_3$O$_7$, where the red symbols are for different R across the lanthanide series when $ x = 0 $ while blue symbols are for increasing $ x $ when R=Y.\cite{mallett2013} The systematic trend reflects the progressive increase in ion size and it is part of a more general trend, shown by the green symbols, across all cuprates.\cite{tallon1990} By any measure the correlation is exceptionally good. This surely gives some hope that underlying the complex soup of electronic interactions which are present in the cuprates there does exist a common and systematic relationship between $ \tcm $ and structure. This paper aims to identify and clarify the details of this relationship in another model cuprate system, the so-called infinite-layer cuprate ACuO$ _{2} $.

But first we comment further on why, despite the competition of compound electronic order, such a relationship might yet exist. To begin with, charge ordering, whether short- or long-range, seems to be confined to the underdoped region\cite{keimer2015, comin2016} and is responsible for the plateau or dip in $ \tcm $ around a hole concentration, $ p $, of about $ 0.12 $~holes/Cu.\cite{chang2012} It therefore likely has little or no influence on the magnitude of $ \tcmaxm $ on the gross scale of $ \tcm $ values ranging from 7~K to 134~K. And while the pseudogap is already present at optimal doping it closes not far beyond,\cite{tallon2001} where the scale of $ \tcm $ is already set. Typically, with increasing hole doping the value of $ \tcm $ falls just 7\% from its maximal value when the pseudogap closes. Indeed the pseudogap is responsible for the decrease in $ \tcm $ on the underdoped side\cite{loram1994} since the magnitude of the gap amplitude, $ \Delta_{0} $, remains essentially constant across the underdoped regime.\cite{yu2008, tallon2011} Evidently it would be more appropriate to seek a relationship between $ \Delta_0 $ and structure since the magnitude of the order parameter is a more fundamental quantity than $ \tcmaxm $. This is underscored by the fact that in the cuprates strong fluctuations reduce $ \tcm $ significantly below the mean field value (thus accounting for the large observed values of $ 2\Delta_{0}/k_{B}\tcmaxm $).\cite{tallon2011} So we embark on our enterprise with the clear understanding that $ \tcmaxm $ is only a proxy for the overall scale of $ \Delta_{0} $, but it is a consistent proxy little affected by charge ordering, the pseudogap and their associated Fermi surface reconstructions. 

The clear implication of Fig.~\ref{fig:0} is that expanding the lattice by substituting larger ions increases $ \tcmaxm $. And so we ask what would be the equivalent effect in the ideal infinite layer compounds ACuO$ _{2} $ with increasing ion size as A progresses down column 2 of the periodic table:  A$ =$ Mg $ \rightarrow $ Ca $\rightarrow $ Sr $ \rightarrow $ Ba? To anticipate, we use the energy-minimized structural parameters found below to calculate $ V_{+} $ for this series. The simplicity of the structure is such that only the $ a $- and $ c $-axis lattice parameters are needed for the calculation. The calculated values of $ V_{+} $ are comparatively high and the black symbols in Fig.~\ref{fig:0} show the model $ \tcmaxm $ values if the correlation is preserved in this idealized system. The predicted $ \tcmaxm $, ranging up to 160~K for A=Ba represent our first key result. The challenge is to synthesise these materials and dope them up to optimal doping $ p \approx 0.16 $ holes/Cu, but this prediction provides strong motivation for deeper study of this generic system. As a comparison, we do have one data point (orange) for (Ca$ _{0.3} $Sr$ _{0.7} $)$ _{0.9} $CuO$ _{2} $ where $ \tcm $ is reported at 110~K.\cite{azuma1992, karpinski1994, karpinski1997} The stoichiometry here suggests a doping level of $ p \approx 0.10 $ and indeed we can estimate the doping level using a second BVS parameter $ V_{-} = 2 + V_{\mathrm{Cu}} - V_{\mathrm{O(2)}} - V_{\mathrm{O(3)}} $, which we find to take the value 0.098, in agreement with the stoichiometry. So this system is somewhat underdoped and we expect $ \tcmaxm $ to exceed 110~K, perhaps more in keeping with the correlation in Fig.~\ref{fig:0}.

We may view the data of Fig.~\ref{fig:0} as generally indicating a decrease in $ \tcmaxm $ due to internal- or chemical-pressure arising from chemical substitution of smaller ions. These \textit{internal}-pressure effects on $ \tcmaxm $ contradict the universal increase of $ \tcmaxm $ under \textit{external}-pressure that is observed in the cuprates.\cite{schillingchapter} This presents a central paradox in cuprate physics and its resolution may well reveal the underlying pairing mechanism. It is difficult to understand in a simple magnetic pairing scenario where both internal and external pressure equally increase the magnitude of the anti-ferromagnetic exchange energy, $ J $. Instead we have argued that ion polarizabilities play a key role in resolving this paradox in an alternative dielectric pairing scenario.\cite{mallett2013} Of course both might be operative. 

This sets the scene for the present studies. There is a clear correlation between structural parameters and $ \tcmaxm $ which can be tuned by changing ion size. On the other hand, internal pressure arising from ion-size effects has the opposite effect of external pressure. We choose the infinite-layer cuprate, ACuO$_2$, as the ideal model system to explore, and possibly reconcile, these effects. We use density functional theory (DFT) calculations where internal-pressure is implemented by altering the A-site ion from Mg to Ba. We then compare these results with the simulated effect of external pressure. ACuO$_2$ was chosen because it has the simplest realisable chemical structure that still displays essential aspects of cuprate physics. It has a P4/mmm tetragonal unit-cell with the Cu ion at $(0,0,0)$, O at $({a}/{2},0,0)$ and $(0,{a}/{2},0)$ and A at $({a}/{2},{a}/{2},{c}/{2})$.  The undoped material is insulating due to strong electronic correlations.\cite{freelon2010} As noted, superconductivity is observed in (Ca$ _{0.3} $Sr$ _{0.7} $)$ _{0.9} $CuO$ _{2} $ with $ \tcm = 110 $ K for hole-doping (due to A vacancies) \cite{azuma1992, karpinski1994} or $ \tcm = 42 $ K under electron-doping (by R substitution for A).\cite{jung2001} More recently, high-quality multilayer thin films of CaCuO$ _{2} $ and SrTiO$ _{3} $ have been grown with $ \tcm $ up to 50~K\cite{dicastro2012, minola2012, salvato2013} where the CuO$ _{2} $ layer is hole-doped via excess oxygen at the SrTiO$ _{3} $-CaCuO$ _{2} $ interface\cite{aruta2013, dicastro2015} and CaCuO$ _{2} $-(Ba$ _{1-x} $Nd$ _{x} $)CuO$ _{2+\delta} $ multilayers have been grown with $ \tcm $ up to 80~K.\cite{balestrino1998, decaro1999, balestrino2003, orgiani2007} 

DFT has been used extensively to study the electronic properties of the cuprates.  Early calculations accurately reproduced phonon spectra and intensities observable by Raman spectroscopy.\cite{cardona1999}  More recently Pavarini  \textit{et al.}\cite{pavarini2001} and Sakakibara \textit{et al.} \cite{sakakibara2012, sakakibara2014} used DFT methods to identify electronic parameters that might play a role in governing $\tcmaxm$. As a model cuprate there have been numerous DFT studies on CaCuO$_2$.\cite{anisimov1991, hatta1992, agrawal1993, andersen1996, massidda1997, wu1999, singh2010}  These studies found that the experimental lattice parameters were reproduced to within a few percent.  The calculated band-structure had similar features to other cuprates, such as YBa$ _{2} $Cu$ _{3} $O$ _{7-\delta} $. Other than a shift in the Fermi-Energy, $ E_{F} $, the band-structure changed little with hole-doping, $ p $, up to 10\% (Ref.~\onlinecite{andersen1996}) - a so-called rigid band shift.  In the DOS there is a van-Hove-singularity feature $ \sim 1 $~eV below $ E_F $ and its position is pressure dependent.\cite{agrawal1995} In addition, various methods have been employed to try to reproduce strong-correlation features of the undoped cuprates, such as the charge-transfer band-gap and the anti-ferromagnetic order on the CuO$ _{2} $ plane.\cite{anisimov1991, wu1999} 
There have also been several reported DFT studies of SrCuO$_2$.\cite{nagasako1997, wu1999} Prepared under ambient pressure, stoichiometric SrCuO$_2$ forms 1D chains \cite{kim1996} whilst the electron-doped Sr$ _{0.9} $R$ _{0.1} $CuO$ _{2} $ forms the P4/mmm tetragonal structure.\cite{jung2001} A recent DFT study investigated the propensity of ACuO$ _{2} $ thin-films lattice matched to their SrTiO$ _{3} $ substrate to form an alternative chain-like structure.\cite{zhong2012} 

In distinction to these previous studies, we present here a comparative study of ACuO$_2$ across the alkaline earth series to investigate internal- and external-pressure effects on the electronic properties in this family of cuprate materials. 

\section{Computational details}
The DFT calculations were performed using the Vienna Ab-Initio Simulation Package \cite{vasp, kresse1993, kresse1994, kresse1996, kresse1996a} (VASP) with Projector Augmented Wave (PAW) pseudopotentials from the VASP5.2 library.\cite{bloch1994, kresse1999}  We used the GGA-PW91 Generalised Gradient Approximation scheme developed by Perdew \textit{et al.} \cite{perdew1992, perdew1993} to derive the form of the exchange-correlation potential and kinetic energy in the single-electron Hamiltonian.

\begin{figure}

\begin{flushleft}
	a)
\end{flushleft} \vspace{-0.6cm}
\includegraphics[width=1.0\columnwidth]{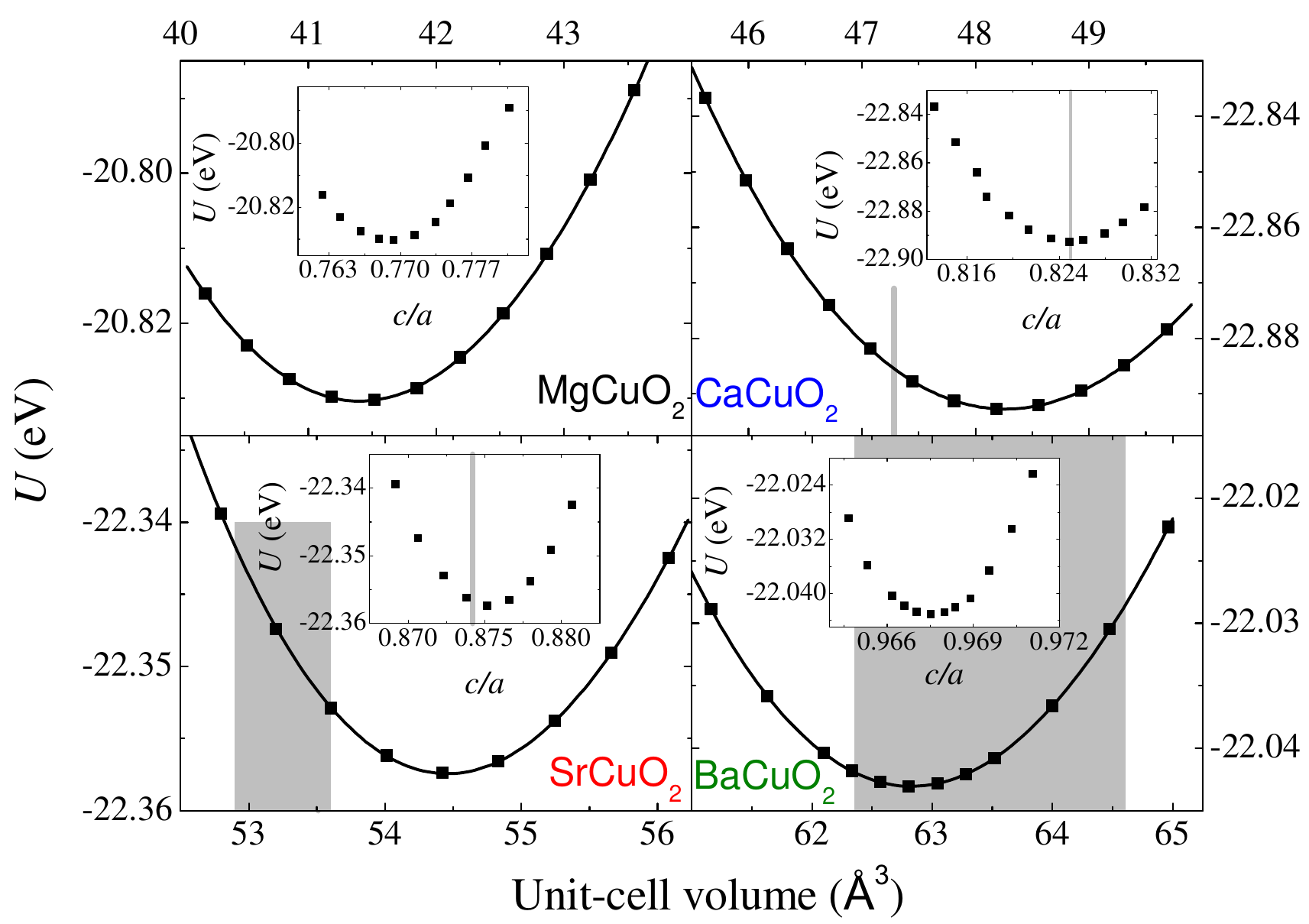}
\begin{flushleft}
	b)
\end{flushleft}
\includegraphics[width=0.9\columnwidth]{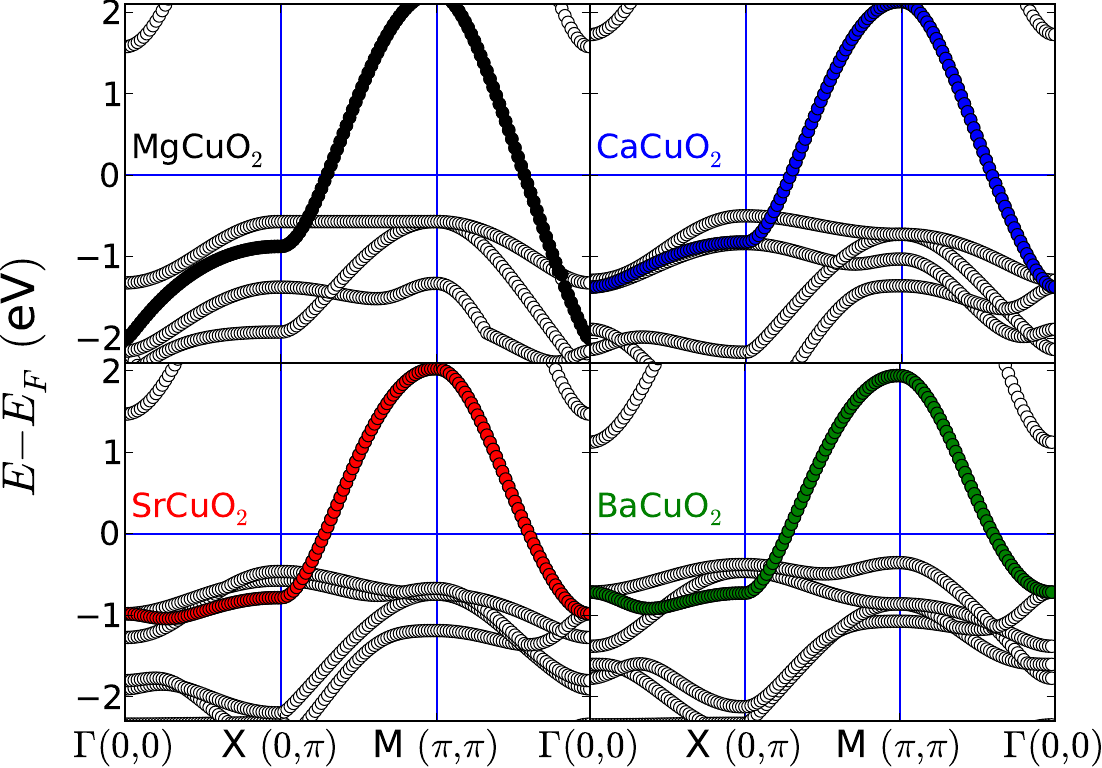}
\begin{flushleft}
	c)
\end{flushleft}
\includegraphics[width=0.88\columnwidth]{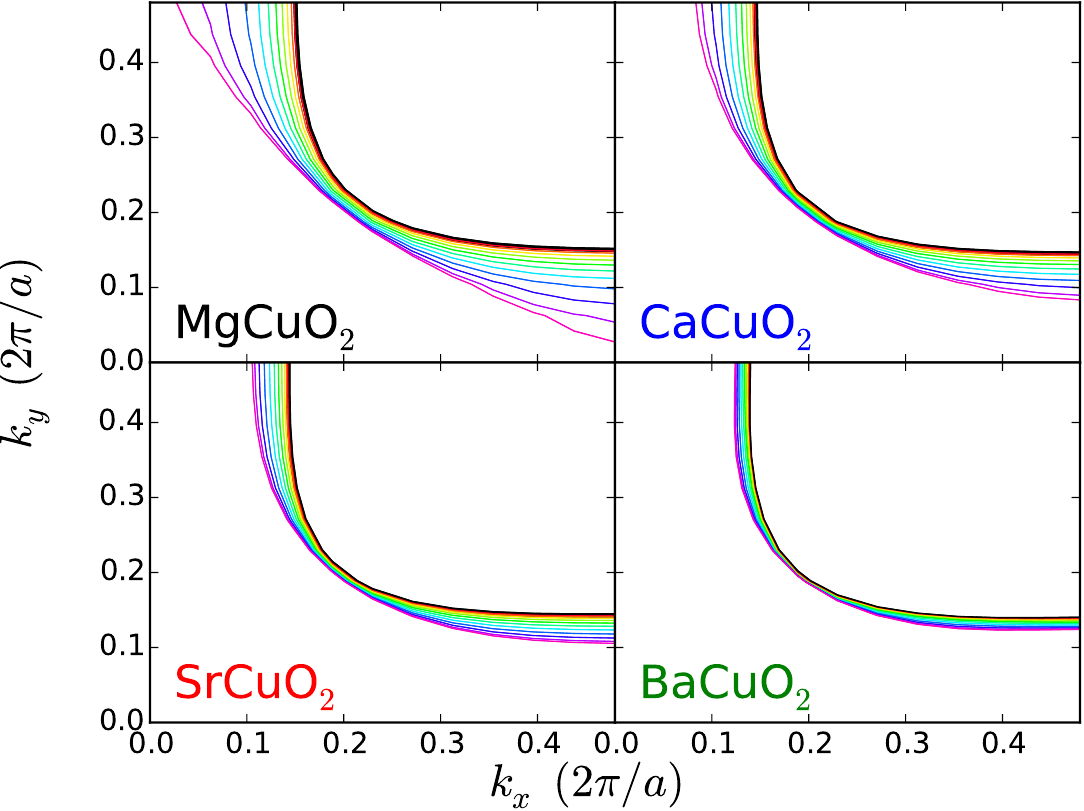}\hspace{0.5cm}

	\caption{\label{fig:1}(Color online) (a) The internal energy, $U$, determined from fixed-unit-cell-volume, ion relaxation calculations on ACuO$_2$ as indicated on each panel.  Insets show the corresponding ${c}/{a}$ lattice parameter ratios. The grey shaded regions represent reported experimental values. The curves in the main panels are fits to the Murnaghan equation of state used to determine the bulk moduli. (b) The band structure for each A along the path $\Gamma$-$X$-$M$-$\Gamma$ with $k_z=0$. (c)  Fermi-contours ($ \ekm =0 $) calculated at $k_{z}$ from 0 (heavy black line) to $ \pi/c $ (magneta line) in steps of $ 0.083 \pi/c $, for each ACuO$ _{2} $.}

\end{figure}

To calculate the equilibrium structural parameters for ACuO$ _{2} $, we determined the internal-energy, $U$, (the `free-energy' in VASP nomenclature) for fixed unit-cell volume, $V$, with a $ 16\times 16\times 16 $ $ \kk $-space, $\Gamma$-centred mesh. For each $V$, the ion positions and lattice parameters were first relaxed, using a conjugate-gradient algorithm \cite{flannery1986} in VASP, and this was followed by an accurate calculation of $U$ with fixed ion positions and lattice parameters.  The results of this procedure are shown in \fig~\ref{fig:1}a). The lattice parameters with the lowest free-energy are; $ a=b=\{3.779$, 3.880, 3.962, $ 4.019\} $~\AA~ and $ c=\{2.907$, 3.201, 3.467, $ 3.888\} $~\AA~ for A=Mg, Ca, Sr and Ba respectively. These correspond to unit-cell volumes of $ V=\{41.52$, 48.19, 54.42, $62.81\} $~\AA$ ^{3} $. The $ c/a $ ratio scales with unit-cell volume as $ c/a \approx \alpha V +const.$ with $ \alpha = 0.0044 $~\AA$^{-3} $, which is similar to what is found experimentally \cite{qin2005structural}. We do not observe any buckling of the CuO$ _{2} $ layers, as expected on symmetry grounds. 

For calculations of the electronic dispersion we use structures with the lowest $ F $ and a $ 24\times 24\times 24$ $ \kk $-space mesh.  Band-structure calculations involve a non-self-consistent calculation using a high-quality self-consistently calculated charge-density as an input.

The grey shaded regions in \fig~\ref{fig:1}a) indicate experimentally determined structural parameters at room-temperature where they are available.\cite{takano1989, karpinski1994, decaro1999, jung2001, qin2005} To our knowledge there are no reports that MgCuO$_2$ has been synthesized.  
Neither has the synthesis of pure bulk BaCuO$_2$ been reported, suggesting this compound is also unstable. For example de Caro \textit{et al.} \cite{decaro1999} report up to 30\% vacancies on the Cu site with lattice parameters $c\approx 4.2$~\AA~and $a=3.906$~\AA~assumed from lattice matching to the SrTiO$_3$ substrate. Here we find a reversed $c/a$ ratio compared with these experimental results. Super-lattices of intercalated Ba-, and Ca- infinite layer compounds have been reported to be superconducting up to 80~K \cite{decaro1999, balestrino2003, orgiani2007}. Note that the ion-size dependent $c/a$ lattice-parameter ratio indicates anisotropic stress under ion-substitution so that the internal-pressure effect is not simply hydrostatic.

To calculate the bulk moduli, $B_0$, we fit the data presented in \fig~\ref{fig:1}a) to the Murnaghan equation of state \cite{khosroabadi2004} (in SI units);
\begin{equation}
F(V)=F_0+ \frac{B_{0}V}{B_{0}'(B_{0}-1)} \left[ B_{0}'(1-V_{0}/V) + (V_{0}/V)^{B_{0}'}-1 \right]  
\end{equation}
 
\noindent where $ B_{0}' = -\partial B_{0}/\partial P |_{T} $ is generally taken to be constant. The fits are shown as lines in \fig~\ref{fig:1}a) from which we find $B_0=98$, 106, 111 and 127 GPa for A = Ba, Sr, Ca and Mg respectively. As expected, these materials become more compressible as the ion-size increases.  By comparison Qin \textit{et al.} \cite{qin2005structural} experimentally determine the bulk moduli of CaCuO$_2$ to be $96$~GPa for $P<6$~GPa and $186$~GPa for $P>6$~GPa, where $ P $ denotes (external) pressure.  (For comparison with other cuprates, $B_{0}=123  $~GPa for YBa$ _{2} $Cu$ _{3} $O$ _{7} $ and $ 78 $~GPa for YBa$ _{2} $Cu$ _{3} $O$ _{6} $.\cite{jorgensen1990}) We thus find reasonable agreement between our calculated stable structural parameters and experimental results. 

To simulate external pressure we used relaxed structures at fixed $ V $ where $ F $ is not at the minimum.  For example, CaCuO$ _{2} $ with a unit-cell volume of $ V= 45.6 $~\AA$^3 $ corresponds to an external pressure of $ P=9.6 $~GPa when using the experimental bulk modulus of 186~GPa (Ref.~\onlinecite{qin2005structural}) and has the ratio $ c/a = 0.813 $. This simulated external pressure has a similar effect on the $ c/a $ ratio as that of internal pressure.  

\section{Results - electronic structure} 
Here we are interested in the internal-pressure induced modification of the electronic dispersion, $\ekm$, close to $E_F$ ($ \ekm = 0 $) and how this compares with that caused by external pressure. 

In \fig~\ref{fig:1}b) we show the band-structure along the path $\Gamma$-$X$-$M$-$\Gamma$ with $k_z=0$ for ACuO$ _{2} $ at $ P=0 $.  Between $\kk=(k_x,k_y)=(0,\pi)$ and $\kk=(\pi,\pi)$ there is a strongly dispersive band crossing $ E_F $ close to $\kk=({\pi}/{4},\pi)$ and $\kk=({\pi}/{2},{\pi}/{2})$.  This band, shown colored, is associated with states in the CuO$ _{2} $ plane and is the only band that contributes to the DOS within $ 0.45 $~eV of $ E_F $.  Between $ -1 $~eV and $ -0.5 $~eV there are several other bands that move closer to $ E_F $ as the A-site ion-size increases (and as $ V $ concurrently increases).

It is well known that the undoped cuprates, including CaCuO$_2$, have a charge transfer gap of approximately 2~eV,\cite{lee2006, freelon2010} in contrast to these calculations which imply undoped ACuO$_2$ is metallic. In fact, this is to be expected from GGA calculations on the undoped system because they do not account for the strong-correlation physics of the undoped cuprates. This deficiency could be addressed by the use of LSDA+U calculations to correct the band gap by the introduction of a Coulomb repulsion term. However, this would cause some difficulties in comparison of ion-size effects, since $ U $ itself should be ion-size dependent due to the effect of ionic polarizability \cite{shannon1993} on screening.\cite{vandenbrink1995, sawatzky2009}

However, the dispersion reflects the rigidly shifted band structure observed using angle-resolved photoemission spectroscopy (ARPES) at finite doping where the strong-correlations are screened out by mobile carriers. Further, the saddle-point vHs is known to reside below $E_F$ and is crossed in the overdoped region leading to a change in the Fermi-surface topology.\cite{kaminski2006, storey2007}  So these calculations do reveal the features of the dispersion that are known to exist when correlations are suppressed and it is therefore plausible that the systematic changes with ion-size shown here reflect real band-structure evolution. 

The calculated band structure also reveals that the dimensionality of the dispersion is affected by internal- and external-pressure.  To illustrate this, in \fig~\ref{fig:1}c we plot Fermi-contours ($\ekm = 0$ in the ($ (k_{x},k_{y} )$-plane) for $k_z$ momenta from 0 to $ \pi/c $ in steps of $ 0.083 \pi/c $. The $k_z$ dispersion is smaller for larger A ion-size because these decouple the CuO$_2$ layers resulting in the more 2-dimensional-like Fermi-surface.  When A=Ba there is little change in the Fermi-contours for different $k_z$. Only when this $k_z$ dispersion is considered is the Luttinger sum rule satisfied for these undoped materials.

\begin{figure}
	\centering
		\includegraphics[width=0.9\columnwidth]{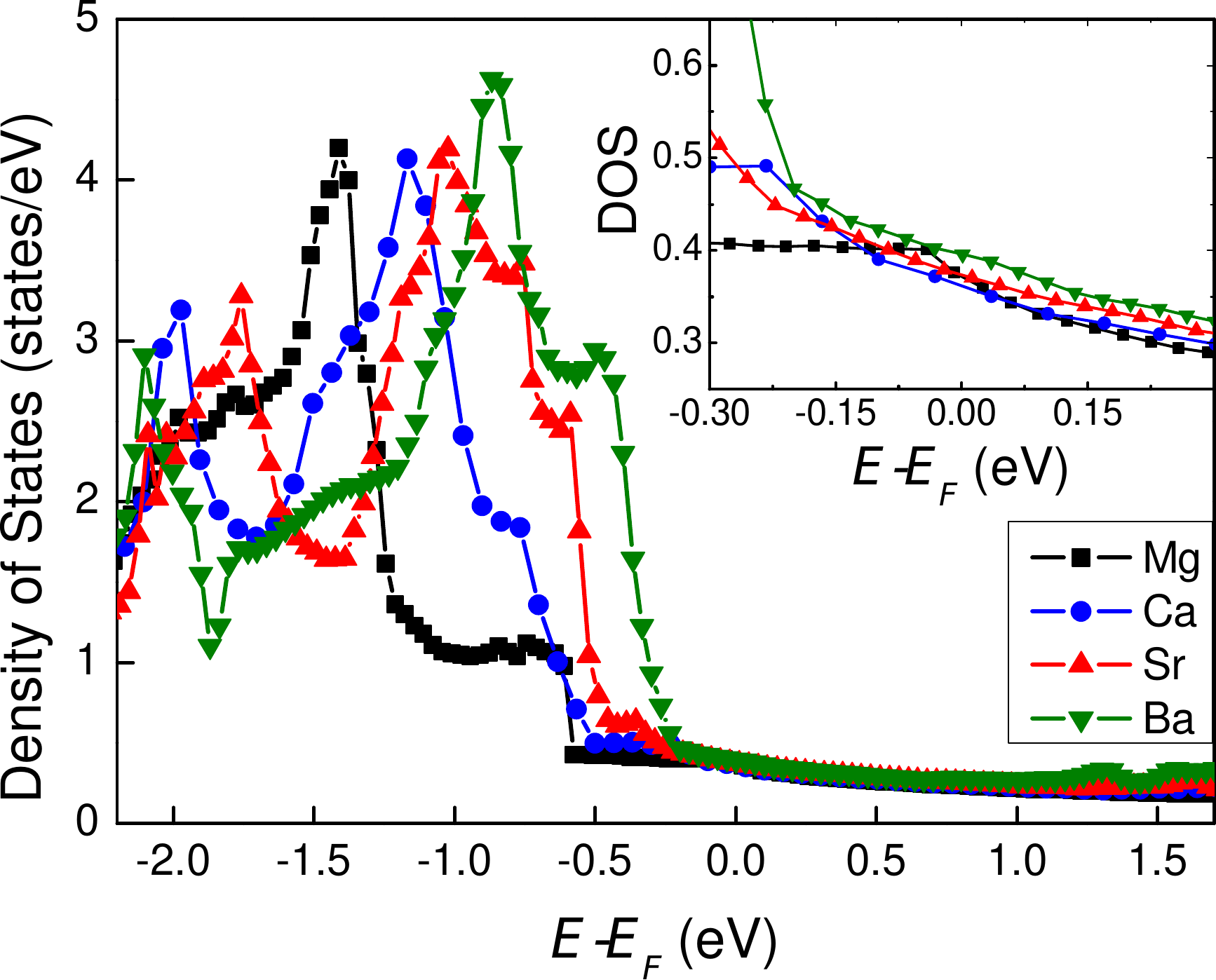}
	\caption{\label{fig:dos}(Color online) The calculated density of states (DOS) for ACuO$ _{2} $ illustrating a shift in weight closer to $E_F$ as the A ion-size increases. The DOS is normalised such that each band contains 1 state.}
\end{figure}

In \fig~\ref{fig:dos} we show the total DOS as a function of internal-pressure (i.e. as the A-site ion-size progressively increases). Starting at $ E=-1.5 $~eV, there is a peak in the DOS that progressively moves closer to $ E_F $ as the internal-pressure increases. The multiple bands in this energy region and the significant $ k_z $ dispersion (particularly for the materials with small unit-cell volume, $ V $) complicate the interpretation of this feature.  There is a secondary peak in the DOS between $ -1 $~eV for MgCuO$ _{2} $ and $ -0.5 $~eV for BaCuO$ _{2} $ that derives in part from the saddle-point in the dispersion of the colored band in \fig~\ref{fig:1}b.  In materials with smaller $ V $ this peak is significantly broadened by the large $ k_z $ dispersion. 

However, the states directly relevant to superconductivity most likely lie within $ \approx 0.5 $~eV of $ E_F $.  This region of the DOS is shown in the inset to \fig~\ref{fig:dos} and shows that the DOS at $ E_{F} $ is similar, but with a slight increase in the DOS with increasing ion size. The band contributing to the DOS in this region, shown colored in \fig~\ref{fig:1}b, can be interpolated using the tight-binding expression;

\begin{equation}
\begin{aligned} 
\ekm = c_0 &+ \tfrac{1}{2} c_1 [\cos(k_x) + \cos(k_y)] \\
&+ c_2 [(\cos(k_x)\cos(k_y)] \\
&+ \tfrac{1}{2} c_3 [\cos(2k_x) + \cos(2k_y)] \\
&+ c_{\perp} \cos(k_z/2)[\cos(k_x)-\cos(k_y)]^2 \\
\end{aligned}
\label{eq:tightbinding}
\end{equation}

\begin{figure}
	\centering
		\includegraphics[width=0.9\columnwidth]{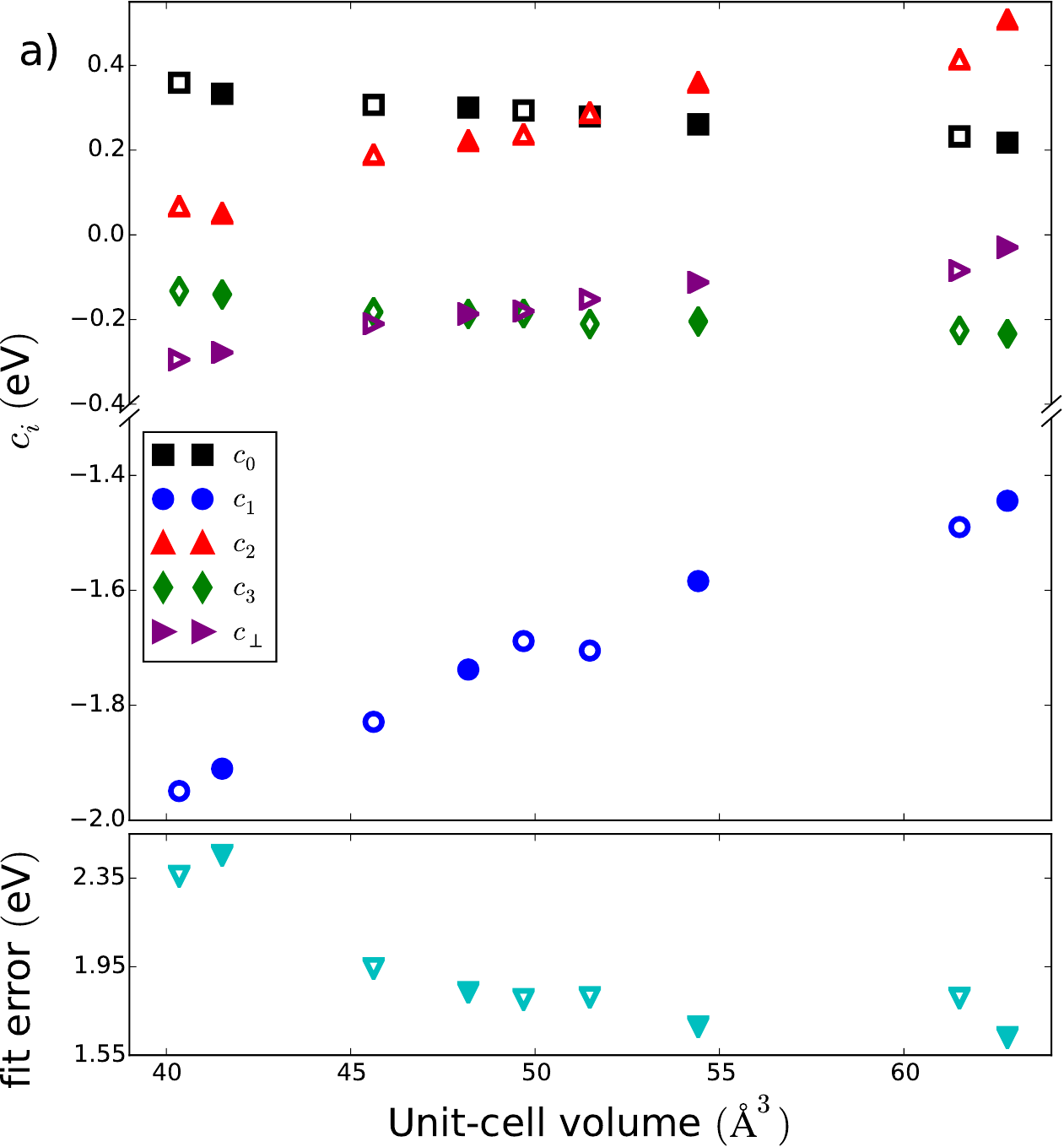}
		\includegraphics[width=0.9\columnwidth]{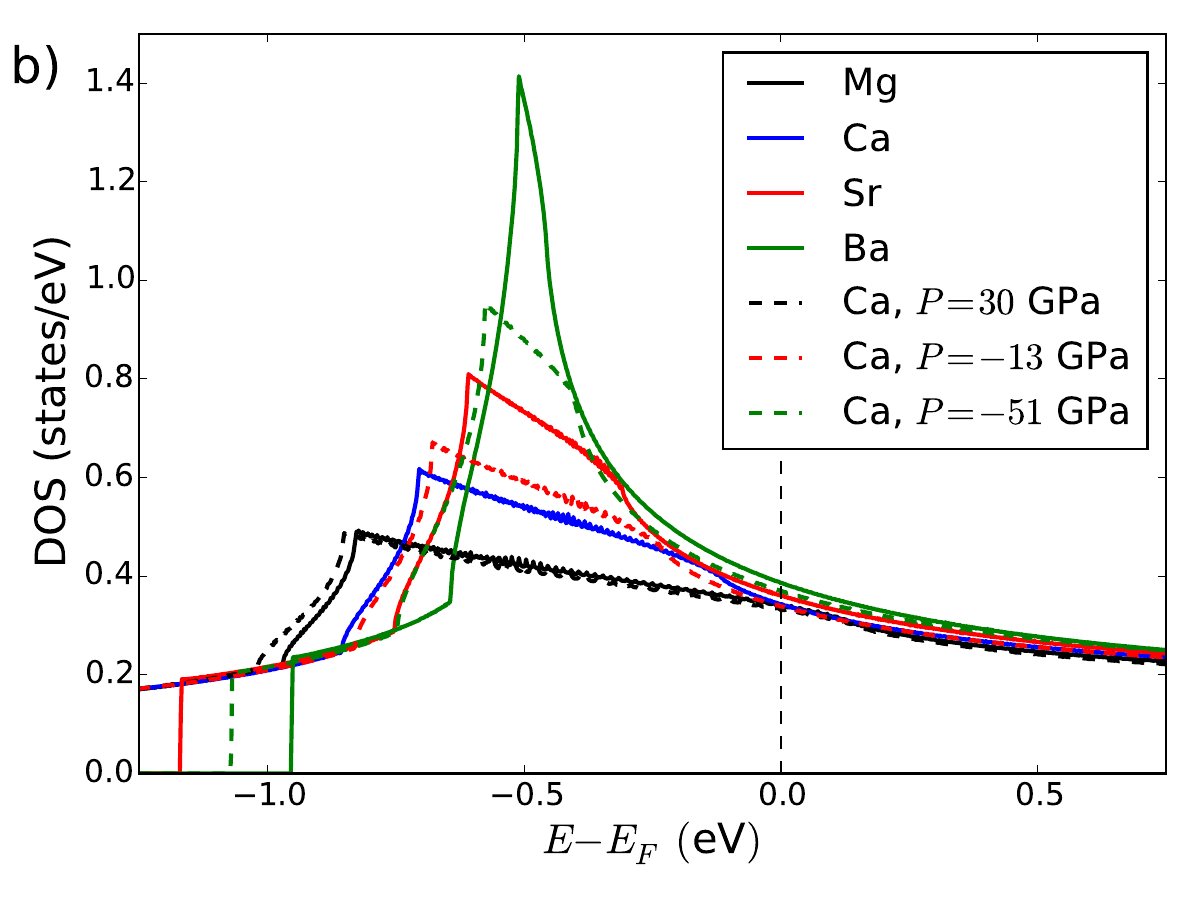}
	\caption{\label{fig:tbparams}(Color online) a) Tight-binding parameters of Eq.~\ref{eq:tightbinding} determined from our DFT calculations for ACuO$ _{2} $ and the fit errors.  Closed (open) symbols are for internal- (external-) pressure effects. 
	b) The density of states (DOS) calculated from the fitted tight binding dispersion (Eq.~\ref{eq:tightbinding}) with the parameters shown in a).  Solid lines represent internal-pressure effects, dashed lines represent external-pressure. The DOS is normalised such that the band contains 1 state. 	}
\end{figure}

\noindent The $ c_{\perp}\cos(k_z/2)[\cos(k_x)-\cos(k_y)]^2 $ term in Eq.~\ref{eq:tightbinding} accounts for the $ k_z $ dispersion \cite{markiewicz2005, pavarini2001} illustrated in \fig~\ref{fig:1}c. The  tight-binding coefficients, $ c_i $, extracted from fitting the DFT-derived dispersion, $ \ekm_{\textnormal{DFT}} $, are plotted in \fig~\ref{fig:tbparams}a. Closed (open) symbols are for internal- (external-) pressure effects and show that the dispersion, $ \ekm $, alters with internal- and external-pressure in a similar fashion. The errors for each tight-binding fit, defined as $N^{-2} \sum_{\kk}{ \sqrt{ \ekm_{\textnormal{DFT}}^2 - \ekm_{\textnormal{fit}} ^2}} $ where $ N $ is the number of data points, are shown in the lower panel.  Typical errors are $ \approx 1.75 $~eV with larger volumes better described by the tight-binding dispersion of Eq.~\ref{eq:tightbinding}.  Eq.~\ref{eq:tightbinding} best describes the band close to $ E_F $ and deviates most from the DFT values around the top and bottom of the band. Parameterizing the dispersion thus is justified if one's interest is restricted to within $ \sim 0.5 $~eV of the $ E_F $, as is indeed the case for our calculations of the superconducting energy gap (described below).  Importantly, this allows us to sample the dispersion over a much finer $ \kk $-space mesh of $ 3000\times 3000 \times 100 $ points (which is not practically possible using VASP alone) for calculations of the DOS and the superconducting energy gap.

The nearest-neighbor hopping term, $ c_1 $, shows the largest pressure-induced variation, increasing from $ -1.9 $ to $ -1.5 $~eV from the smallest to largest $ V $.  Concurrently, the next-nearest-neighbor term, $ c_2, $ increases from $ 0.05 $ to $ 0.5 $~eV resulting in the more `rounded' Fermi-contours of BaCuO$ _{2} $ seen in \fig~\ref{fig:1}c. We describe later in the paper how these pressure-induced changes in Fermi-surface shape relate to the superconducting energy gap.  $ c_{\perp} $ vanishes as $ V $ increases reflecting the weaker $ c $-axis coupling between CuO$ _{2} $ planes at larger $ V $ and a more 2-dimensional-like Fermi-surface.

The DOS calculated from the parameterized dispersion, $ \ekm_{\textnormal{fit}} $, is shown in \fig~\ref{fig:tbparams}b.  Full solid lines represent internal-pressure effects (ACuO$ _{2} $ with A indicated in the legend) and dashed lines represent external-pressure effects (CaCuO$ _{2} $ with the effective pressure indicated in the legend). \fig~\ref{fig:tbparams}b reveals similar features to the DOS calculated with VASP where materials with larger $ V $ have higher DOS at $ E_{F} $. \fig~\ref{fig:tbparams}b also shows the vHs at $ E\approx -0.5 $~eV originating from the saddle-point in the band around $ (k_x,k_y)=(0,\pi) $ (and symmetry related points).  The vHs is significantly broadened by the $ k_z $ dispersion for materials with smaller $ V $.

\section{Superconducting energy gap calculations}

What bearing might these internal- and external-pressure induced changes in the dispersion and DOS have on superconductivity in these systems?  To approach this question we solve the self-consistent BCS gap-equation \cite{zhou1992, storey2007} to obtain the superconducting energy gap, $ \Delta_0 $; 

\begin{equation}
\Delta(\kk) = -\frac{1}{2}\sum_{\kk^{\prime}}{ \frac{ v_{\kk\kk^{\prime}} \Delta(\kk^{\prime})}{ \sqrt{\ekpm^2 + \Delta(\kk^{\prime}) ^2} } }
\label{eq:gapeq}
\end{equation}

\noindent Here $ v_{\kk\kk'} $ is the pairing potential of the form $v_{\kk\kk'}= vg_{\kk}g_{\kk'} $ where $ g_{\kk} = \cos(k_x) - \cos(k_y)$ to reflect $ d $-wave symmetry, and the $ d $-wave superconducting gap is given by $ \Delta(\kk)= \frac{1}{2}\Delta_0 g_{\kk} $. Note that in the absence of competing order parameters, such as the pseudogap, one can expect that $ \tcm $ is proportional to $ \Delta_0 $. We take $ \ekm $ from the tight-binding fits, $ \ekm_{\textnormal{fit}} $, discussed above.
 
\begin{figure}
	\centering
	\includegraphics[width=0.9\columnwidth]{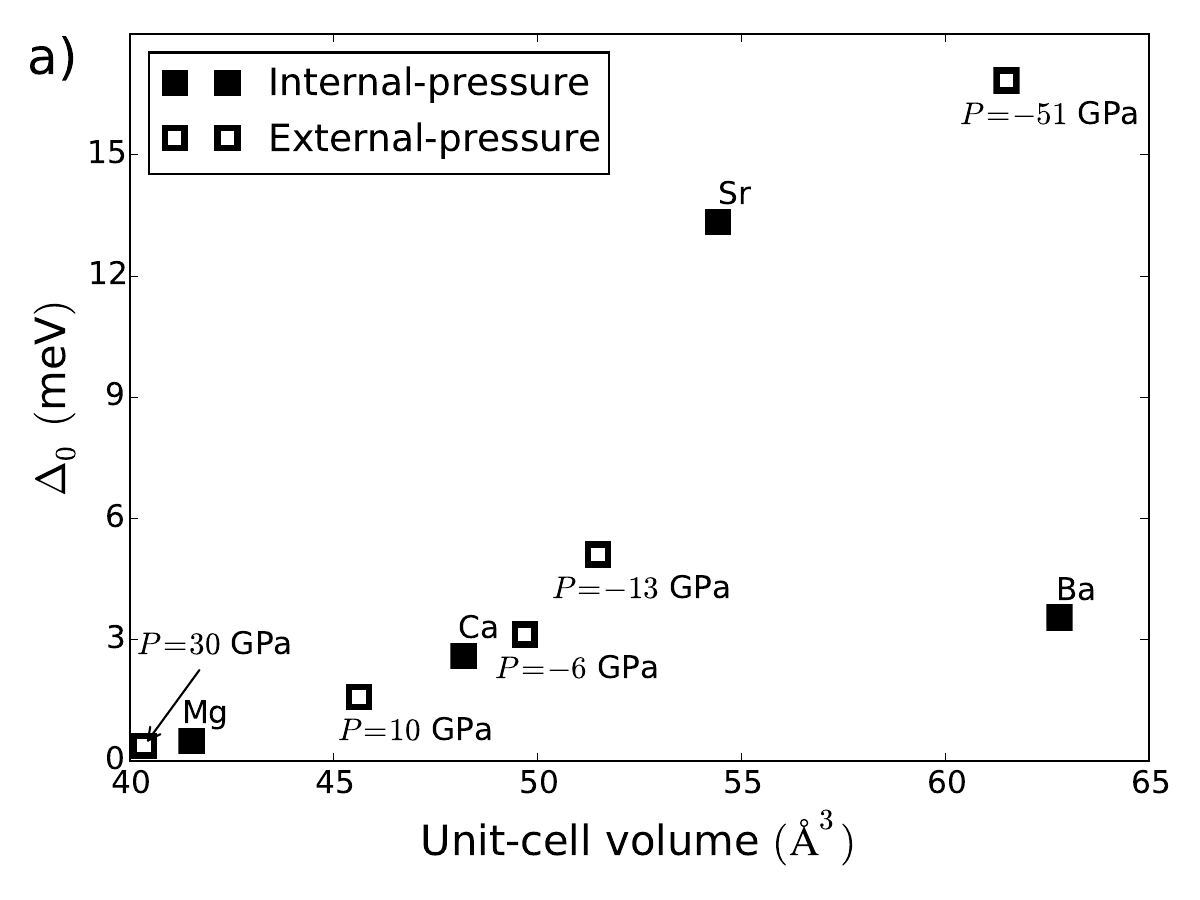}
	\includegraphics[width=0.9\columnwidth]{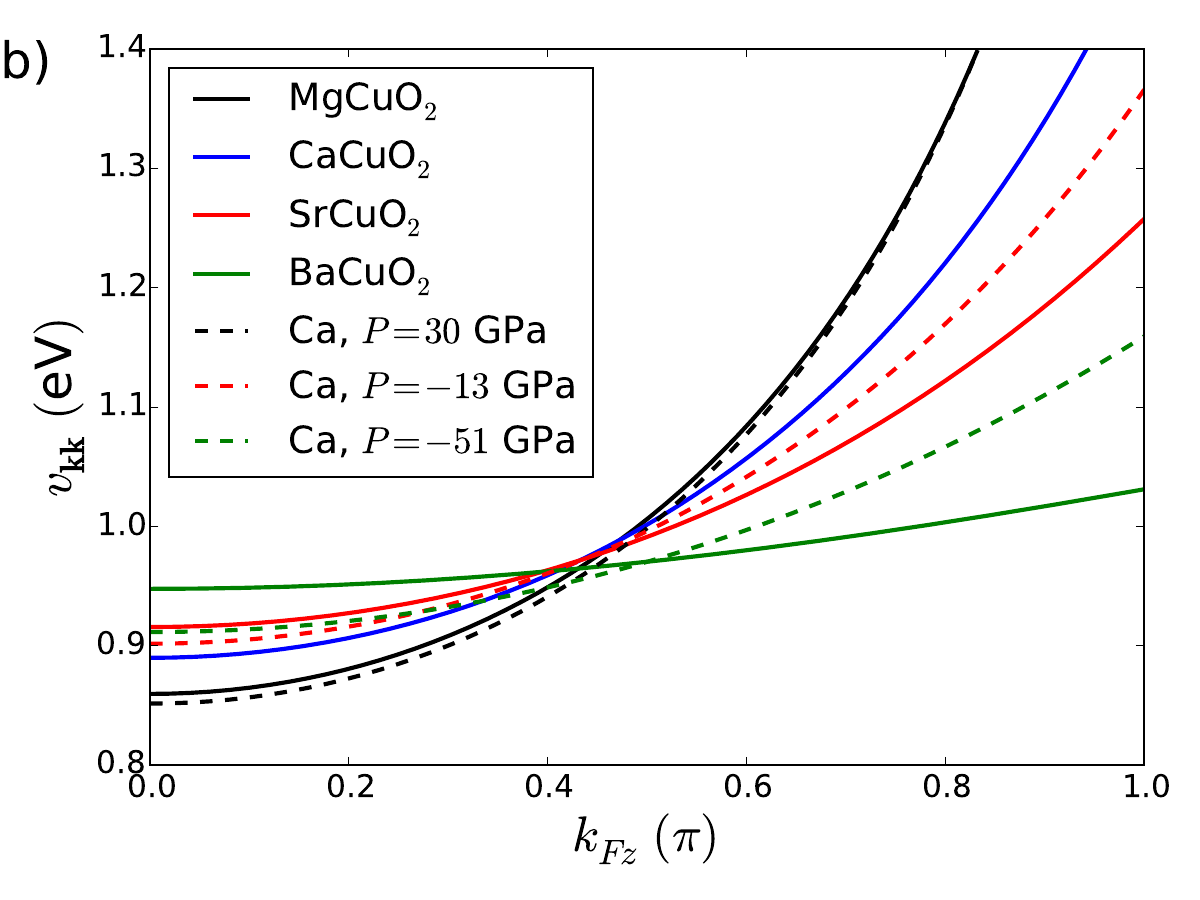}
	
	\caption{\label{fig:gaps} a) The superconducting energy gap, $ \Delta_0 $, determined from the self-consistent BCS gap equation, Eq.~\ref{eq:gapeq}, as a function of the unit-cell volume.  Solid symbols represent internal-pressure induced by A-site ion substitution in ACuO$ _{2} $ and open symbols represent CaCuO$ _{2} $ under positive and negative external-pressure. There is a significant variation in $ \Delta_0 $ that is correlated with the unit-cell volume somewhat independent of how the pressure is applied. b) The magnitude of the assumed pairing interaction, $ v_{\kk\kk} $, corresponding to the points where the Fermi contours touch the zone boundary at $ (k_x,k_y) = (\pi, k_{Fy}) $ as a function of $ k_{Fz} $.  The Fermi-surface of BaCuO$ _{2} $ spans regions of lower pairing potential than SrCuO$ _{2} $ which results in the comparatively small $ \Delta_0 $ in a).	
		}
\end{figure}

We solve Eq.~\ref{eq:gapeq} self-consistently assuming a pressure and ion-size independent $ v=350 $~meV and summing over states $ \ekm $ within $ \pm 150 $~meV of $ E_{F} $.  The resulting $ \Delta_0 $ values are plotted as a function of unit-cell volume in \fig~\ref{fig:gaps}.
Closed symbols represent volume changes induced by internal-pressure, while the open symbols represent volume changes induced by external-pressure. There is a 30-fold increase in the magnitude of $ \Delta_0 $ for an approximately $ 50 $\% increase in the unit-cell volume.  The sensitivity of $ \Delta_0 $ on the unit-cell volume shown in these calculations demonstrates that these internal- and external-pressure effects can potentially have a dominant effect on the superconducting properties via the electronic dispersion, $ \ekm $.

Although there is some scatter in the internal-pressure data which ultimately results from the detailed shape of the Fermi-surface, it appears that $ \Delta_0 $ correlates with the unit-cell volume independent of the mechanism altering the volume.  This means that the character of the band crossing $ E_F $ derives from the CuO$ _{2} $-layer and not the A-site ion.  This correlation is independent of the choice of $ v $ or energy cut-off. We also checked this correlation for different doping levels of the CuO$ _{2} $-layer, $ p $, where $ p $ is varied by setting $E_F $ so that integrating the DOS (derived from Eq.~\ref{eq:tightbinding}) up to $ E_F $ gives the desired number of electrons, e.g. $ 0.84 $ if $ p= 0.16 $ is desired. This method of varying $ p $ can only be valid within a rigid-band scenario. At $ p=0.19 $, $ E_F $ is shifted down in energy by up to 280~meV in \fig~\ref{fig:1}b so that there is still only one relevant band for the calculation of $ \Delta_0 $.  With $ p>0 $ we find a similar correlation between $ \Delta_0 $ and the unit-cell volume yet more clearly demonstrated.  

We note that the relative magnitudes of $ \Delta_0 $ for SrCuO$ _{2} $ and BaCuO$ _{2} $ (filled squares at $ V=55 $ and $ 61.5 $~\AA$ ^{3} $ respectively) are at odds with the general trend in $ \Delta_0 $ and with CaCuO$ _{2} $ at equivalent $ V $ (open squares) and also at odds with the general trends in \fig~\ref{fig:0}.  Recall that the tight-binding parameters determining $ \ekm $ alter under internal- and external-pressure in a similar fashion so that one might expect similar $ \Delta_0 $ from Eq.~\ref{eq:gapeq}.  Furthermore, the DOS at the Fermi-level, $ N(E_F) $, scales only with $ V $ and so, in light of the weak-coupling BCS result for $ d $-wave superconductivity:

\begin{equation}
\Delta_0 = 4.28k_BT_c = 4\hbar\omega_B \exp \left( \frac{-1}{N(E_F)v} \right)
\end{equation}

\noindent the distinct $ \Delta_0 $ magnitudes for SrCuO$ _{2} $ and BaCuO$ _{2} $ are especially surprising.  In this equation $ \hbar\omega_B $ is the pairing-boson energy scale. However, one can understand these disparate $ \Delta_0 $ as resulting from the difference in the average pairing potential close to the anti-nodes, where $ v_{\kk\kk} $ is largest.  This is in turn due to the detailed shape of the Fermi-surface.  By way of illustration we plot in \fig~\ref{fig:gaps}b $ v_{\kk\kk} $ corresponding to the points where the Fermi contours touch the zone boundary at $ (k_x,k_y) = (\pi, k_{Fy}) $ as a function of $ k_{Fz} $ (we append the subscript $ F $ to denote a value of $ \kk $ at the Fermi-energy). Combined, \fig~\ref{fig:1}c and \fig~\ref{fig:gaps}b show that the Fermi-surface of BaCuO$ _{2} $ spans regions of lower pairing-potential than SrCuO$ _{2} $ which results in the comparatively small $ \Delta_0 $. This result highlights how sensitive $ \Delta_0 $ can be to the precise shape of the Fermi-surface.

\section{Discussion}

While \fig~\ref{fig:gaps}a shows that $ \Delta_0 $ correlates with $ V $ somewhat independently of how the pressure is applied, because of the near linear relationship between $ V $, $ c_1 $ and $ c_2 $ (see \fig~\ref{fig:tbparams}), the same correlation between $ \Delta_0 $ and the composite parameter $ -c_2/c_1 $ or $ (-c_2 + c_3)/c_1 $ holds. The ratio $ r \sim t'/t \equiv -c_2/c_1 $ is the Fermi-surface-shape parameter which Pavarini \textit{et al.}, and more recently Sakakibara \textit{et al.}, showed correlates with $ \tcmaxm $ across a wide range of cuprates \cite{pavarini2001, sakakibara2012, sakakibara2014} and this is consistent with the correlation between $ -c_2/c_1 $ and $ \Delta_0 $ that we find here.

It is interesting to relate this relation between Fermi-surface-shape and $ \Delta_{0} $ or $ \tcmaxm $ to that between $ V+ $ and $ \tcmaxm $ shown in \fig~\ref{fig:0}. The large values of $ V_{+} $ in \fig~\ref{fig:0} which lead to large projected values of $ \tcmaxm $ arise mostly because of the absence of the apical oxygen in the infinite-layer system. It is this which also yields the high $ \tcm $ values of the three-layer cuprates.\cite{haines1992} The physical reason is likely related to the distribution of charge on the (planar) oxygen orbitals relative to the copper\cite{ohta1991}, which was what the parameter $ V_{+} $ was first introduced as a measure of\cite{tallon1990} (we note that our DFT calculations here can confirm that the fraction of charge on oxygen orbitals grows concurrently with $ V_{+} $ in ACuO$ _{2} $).  Such ideas continue to find support in recent experimental work. Peng \textit{et al.} \cite{peng2016} highlight the role of the apical oxygen on the `range' of electronic interactions and $ \tcmaxm $ where they found that, in general, as the apical oxygen is moved away from the CuO$ _{2} $ layer the long-range Heisenberg interaction terms become more important for an accurate description the experimentally measured magnon dispersion. Also, Rybicki \textit{et al.} have recently made a similar prediction for very high $ \tcmaxm $ values in the electron-doped systems based on their NQR measurements of charge distribution between oxygen and copper orbitals.\cite{rybicki2016} Presumably they would find the same for the infinite layer compounds and an important challenge in this regard is doping these compounds. In heterostructure superlattices of these compounds the significantly doped CuO$ _{2} $ layers are found at the interface and are effectively bonded to an apical oxygen that is part of the adjacent layer.\cite{balestrino1998, aruta2013, dicastro2015} This changes the local structure of the infinite-layer compound. Only a modest transfer of charge, over the range $ \sim 5 $~\AA,\cite{dicastro2015} dopes the CuO$ _{2} $ layers beneath the interface, with the innermost layers remaining undoped and non-superconducting or heavily underdoped. It remains to be seen whether alternate approaches to doping the infinite layer compounds, such as liquid ion-gating, are viable.

In conclusion, we draw three main inferences: firstly, we expect from BVS correlations that $ \tcmaxm $ values for optimally-doped infinite-layer compounds ACuO$ _{2} $ are rather high, rising to $ \tcmaxm \approx 160 $~K as A ranges from Mg to Ba (and possibly higher still for A=Ra) due to internal pressure effects. Secondly, we show from DFT calculations that such marked changes in superconducting properties are indeed plausible via the systematic changes in the dispersion, $ \epsilon(\kk) $. Thirdly however, leaving aside Ba, these effects prove to be essentially the same for both internal- and external-pressure in contrast with the opposing effects of internal- and external-pressure on $ \tcm $ observed experimentally, for example in the RBa$ _{2} $Cu$ _{3} $O$ _{7-\delta} $ (Refs.~\onlinecite{guillaume1994, licci1998, gilioli2000, schillingchapter}) and (La$_{1-x}$Ca$_x$)(Ba$_{1.75-x}$La$_{0.25+x}$)Cu$_3$O$_y$ (Refs.~\onlinecite{ofer2006, sanna2009}) systems. This work suggests that the pressure-induced modification of $ \epsilon(\kk) $ alone is insufficient to account for the observed pressure-induced variation in $ \tcmaxm $, and that the disparate effects of internal and external pressure are to be found in the pairing potential, $ v(\kk) $, or the pairing-boson energy-scale. As we showed previously, a dielectric pairing model seems to account for this disparity better than a magnetic model.\cite{mallett2013} Moreover, we find from these calculations that in order to achieve realistic gap values of the order of 30~meV, as observed, one needs a bosonic energy scale of the order of 1~eV, substantially larger than the magnetic energy scale and perhaps more consistent, for example, with a dielectric model where pairing is mediated by exchange of virtual polarisation waves.\cite{atwal2004} Clearly there is a delicate interplay between the structural, electronic and superconducting properties in the cuprates and a deeper understanding of this is necessary to elucidate the dramatic material specificity of $ \tcm $. Complementary internal- and external-pressure effects offer a powerful approach to advance this project.

\begin{acknowledgements}
 The authors would like to thank D. Schebarchov and D. Mollenhauer for helpful discussions. This work was supported by the Marsden Fund of New Zealand.
\end{acknowledgements}

\end{document}